\documentclass[prb, preprint]{revtex4}
\usepackage{graphicx}
\usepackage{tikz,pgfplots}
\usepackage{amsmath,amssymb}
\usepackage[]{subfigure}

\begin{document}
\title{A Pedagogical Extension of the One Dimensional Schr\"odinger's Equation to Symmetric Proximity Effect System Film Sandwiches}
\author{B.J. Luke and P.R. Broussard}
\affiliation{Covenant College, Lookout Mountain, GA 30750}

\begin{abstract}
This study sought to use Schr\"odigner's equation to model superconducting proximity effect systems of symmetric forms. As N. R. Werthamer noted, [Phys. Rev. \textbf{132} (6), 2441 (1963)] one to one analogies between the standard superconducting proximity effect equation and the one-dimensional, time-independent Schr\"odinger's equation can be made, thus allowing one to model the behavior of proximity effect systems of metallic film sandwiches by solving Schr\"odinger's equation. In this project, film systems were modeled by infinite square wells with simple potentials. Schr\"odinger's equation was solved for sandwiches of the form $S(NS)_M$ and $N(SN)_M$, where $S$ and $N$ represent superconducting and nonsuperconducting metal films, respectively, and $M$ is the number of repeated bilayers, or the period. A comparison of Neumann and Dirichlet boundary conditions was done in order to explore their effects. The Dirichlet type produced eigenvalues for $S(NS)_M$ and $N(SN)_M$ sandwiches that converged for increasing $M$, but the Neumann type produced eigenvalues for the same structures that approached two different limits as $M$ increased. This last behavior is unexpected, as it implies a dependence upon the type of film end layer.
\end{abstract}

\maketitle

\section{Introduction}
As discussed in Ref.~\onlinecite{AJP_paper}, the superconducting proximity effect is normally modeled by the de Gennes-Werthamer equation, given by 
\begin{equation} \label{prox eq}
\chi\bigg(-\xi^2\frac{d^2}{dx^2}\bigg)\Delta(x)+\ln\bigg(\frac{\theta_D}{T_c(x)}\bigg)\Delta(x)=\ln\bigg(\frac{\theta_D}{T_c}\bigg)\Delta(x),
\end{equation}
where $\chi(z)=\psi(1/2+z/2)-\psi(1/2)$ (where $\psi(x)$ is the diGamma function), $\Delta(x)$ is the self-energy function, $\xi$ depends on material properties in the superconductor or metal at point $x$, $\theta_D$ is the Debye temperature, $T_c(x)$ is the superconducting transition temperature for the specific material at point $x$ in isolation, and $T_c$ is the superconducting transition temperature of the composite system. In Ref.~\onlinecite{AJP_paper}, analogies were used (which were first made by N. R. Werthamer\cite{Werthamer}) between Eq. \ref{prox eq} and the one dimensional, time-independent Schr\"odinger's equation, given by
\begin{equation} \label{schrod eq}
-\frac{\hbar^2}{2m}\frac{d^2\psi(x)}{dx^2}+V(x)\psi(x)=E\psi(x).
\end{equation}
Here $\psi(x)$ serves conventionally as the wavefunction, and will for the rest of the paper. The structural similarities between Eq. \ref{schrod eq} and Eq. \ref{prox eq} are easy to spot: the wavefunction $\psi(x)$ corresponds to $\Delta(x)$, $\hbar^2/2m$ to $\xi^2$, $V(x)$ to $\ln(\theta_D/T_c(x))$, and $E$ to $\ln(\theta_D/T_c)$. The point is that if we understand the nature of solutions to Schr\"odinger's equation, we can also understand the behavior of proximity effect systems. 

The other benefit about this technique (as Ref.~\onlinecite{AJP_paper} also mentions) lies in its pedagogical applications in the undergraduate classroom. The fact that we are using the one-dimensional Schr\"odinger's equation to solve proximity systems makes the study of these systems accesible to students who have completed introductory quantum mechanics courses. Even students with basic knowledge of quantum theory should be able to make use of this method, as the solutions to Schr\"odinger's equation for infinite square wells (which we will use here) can be fairly easily obtained with only basic knowledge of quantum theory. 

\section{Details of the Problem}
\begin{figure}
\includegraphics*[width=6.0in]{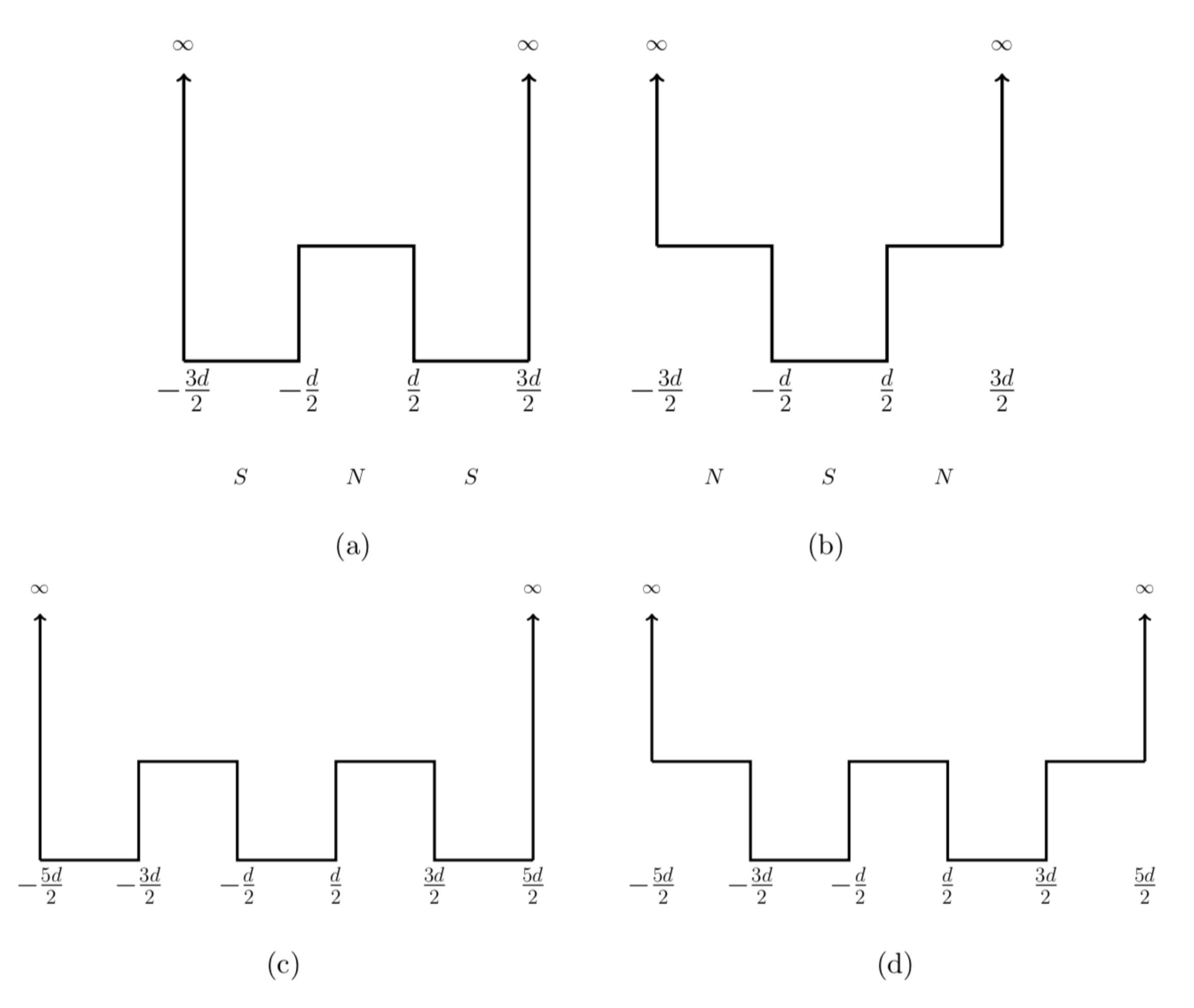}
\caption{Infinite Square Wells with simple potentials modeling one and two period proximity systems for both $S(NS)_M$ and $N(SN)_M$ cases. The potentials here are constant, and all regions of nonzero potentials are $V(x)=V$. While only $M=1,2$ wells are shown here, the same pattern can be trivially extended to greater $M$.}
\label{ISW}
\end{figure}

In order to understand more precisely how we will use Eq. \ref{schrod eq}, we must further describe the details of the problem at hand. To model proximity layer systems of alternating $N$ and $S$ layers, we will use the same model as Ref.~\onlinecite{AJP_paper}: infinite square wells (ISW) with simple constant potentials. Regions of nonzero potential are indicative of nonsuperconducting layers (or layers with a lower transition temperature) since $V$ has an inverse relationship to $T_c(x)$, and, conversely, regions of $V(x)=0$ are representative of superconducting layers. It should also be noted that we do not have to vary the width of the layers directly in order to understand what happens when the layer width gets small. We will be measuring the energies and potentials in units of $\hbar^2/(2md^2)$ (as Ref.~\onlinecite{AJP_paper} did), so an increase in $V$ has the same effect as a decrease in $d$. A sample of the wells that will be used is shown in Fig. \ref{ISW}.

Further, it should be mentioned what will be expected of the wavefunction. Since the wavefunction serves to represent the self-energy function, we must then expect $\psi(x)$ to obey the nature of $\Delta(x)$. The implication is that since $\Delta$ is described by only one space coordinate $x$, then the energy gap $E_0$ of the system is equal to the minimum value of $|\Delta(x)|$.\cite{DeGennes} This means that $\Delta$ cannot go through zero, thus we will demand that $\psi(x)$ be positive definite. Lastly, we will only be interested in solutions where $E<V$ since $T_c$ is always assumed to be between the transition temperatures of the two metals. 

The main purpose of this paper is to explore how boundary conditions have an effect on proximity effect system film sandwiches of the forms $S(NS)_M$ and $N(SN)_M$. Here we will (like Ref. ~\onlinecite{AJP_paper}) compare a Dirichlet and Neumann boundary condition on $\psi(x)$. In actuality, the Neumann condition is what is required for $\Delta(x)$ when using the standard de Gennes-Werthamer approach to solve these systems (i.e., using Eq. \ref{prox eq}).\footnote{The Neumann condition is physically necessary for proximity effect systems, as described by de Gennes.\cite{DeGennes}} However, the theoretical results from the standard solutions defeat what one would intuit about the behavior of such systems. What they show is that the transition temperature of the composite structure $T_c$ depends greatly on the type of end layer that the structure has (i.e., whether the structure is enclosed by $S$ or $N$ film layers), even as the number of layers between the end layers increases.\cite{PRB Paper} What this means is that $T_c$ for $S(NS)_M$ systems and $T_c$ for $N(SN)_M$ systems do not converge as $M\rightarrow\infty$. This, of course, is a counterintuitive result, so here the point is to determine how the type of boundary condition affects the theoretical behavior of $T_c$ for the two different types of film sandwiches.

\section{Solutions to Schr\"odinger's Equation}
We now have the job of finding solutions to Schr\"odinger's equation for the potential wells describing our relevant film sandwiches. We will solve the systems in the symmetric case (i.e., for an even function $\psi(x)$) using linear combinations of sines, cosines, and hyperbolic sines and cosines. The solutions will necessarily be piecewise due to the discontinuous nature of $V(x)$, so in regions of zero potential, the wavefunctions will be of the form $\psi_n(x)=\alpha_n\cos(kx)+\beta_n\sin(kx)$, and in regions of $V(x)=V$ the wavefunctions will be of the form $\psi_n(x)=\alpha_n\cosh(qx)+\beta_n\sinh(qx)$, where $q=\sqrt{2m(V-E)/\hbar^2}$ and $k=\sqrt{2mE/\hbar^2}$. (The index $n$ denotes different layers, with $n=0$ corresponding to the layer across the symmetry point $x=0$ and $n$ increasing and decreasing by integer values to the right and to the left of $x=0$, respectively.) We shall begin by simply solving the case for the $SNS$ (one period) system, modeled in Fig. \ref{ISW}a, under the Dirichlet condition. 

\subsection{Dirichlet Condition}
For simplicity, we will denote the region $-\frac{3d}{2}<x<-\frac{d}{2}$ with a subscript $n=-1$, region $-\frac{d}{2}<x<\frac{d}{2}$ with $n=0$, and region $\frac{d}{2}<x<\frac{3d}{2}$ with $n=1$. However, since the function is symmetric, we need only deal with half of the layers. The Dirichlet condition requires that $\psi=0$ at $x=-\frac{3d}{2}$ and at $x=\frac{3d}{2}$. In addition, the symmetry condition requires that $\psi^\prime(0)=0$ about $x=0$. Hence our equations will be of the form $$\psi_{0}(x)=\alpha_0\cosh(qx),$$ and $$\psi_1(x)=\beta_1\sin\left[k\left(\frac{3d}{2}-x\right)\right].$$ Enforcing the continuity requirement of $\psi$ at the $x=\frac{d}{2}$ interface gives $\alpha_0\cosh\frac{qd}{2} = \beta_1\sin kd$ and enforcing the continuity requirement of $\psi^\prime$ at the same location gives $q\alpha_0\sinh\frac{qd}{2} = -k\beta_1\cos kd$. Dividing these equations yields the eigenvalue equation:
\begin{equation*}
\dfrac{q}{k}\tanh\frac{qd}{2} = -\cot kd.
\end{equation*}

If we now consider the eigenvalue equation for $S(NS)_2$ (Fig.~\ref{ISW}c), it is found in a similar fashion. The relevant wavefunctions for each layer (from $\psi_0$ to $\psi_2$, since the others are superfluous) are $$\psi_0(x) = \alpha_0\cos(kx),$$ $$\psi_1(x)=\alpha_1\cosh\left[q\left(\frac{3d}{2}-x\right)\right]+\beta_1\sinh\left[q\left(\frac{3d}{2}-x\right)\right],$$ and $$\psi_2(x)=\beta_2\sin\left[k\left(\frac{5d}{2}-x\right)\right].$$ By enforcing the same continuity requirements at the internal interfaces, and after much algebra, one arrives at the eigenvalue equation:
\begin{equation*}
\dfrac{k}{q}\tan\frac{kd}{2} = \dfrac{\delta_2\tanh qd + 1}{\delta_2 + \tanh qd}
\end{equation*}
where $\delta_2 = \dfrac{q}{k}\tan kd$.

Using the same methodology for $M=3$, we arrive at an eigenvalue equation given by
\begin{equation*}
\dfrac{q}{k}\tanh\frac{qd}{2} = \dfrac{\delta_3\tan kd - 1}{\delta_3 + \tan kd}
\end{equation*}
where $\delta_3 = \dfrac{k}{q}\bigg[\dfrac{\delta_2 + \tanh qd}{\delta_2\tanh qd + 1}\bigg]$. For $M=4$, we get
\begin{equation*}
\dfrac{k}{q}\tan\frac{kd}{2} = \dfrac{\delta_4\tanh qd + 1}{\delta_4 + \tanh qd}
\end{equation*}
where $\delta_4 = -\dfrac{q}{k}\bigg[\dfrac{\delta_3 + \tan kd}{\delta_3\tan kd - 1}\bigg]$. By this point, it becomes evident that there is a recursive nature to these equations. What we find is that for $M$ even
\begin{equation}
\begin{gathered}
\dfrac{k}{q}\tan\frac{kd}{2} = \dfrac{\delta_M\tanh qd + 1}{\delta_M + \tanh qd},\\
\delta_{M, \mathrm{\ even}} = -\dfrac{q}{k}\left[\dfrac{\delta_{M-1} + \tan kd}{\delta_{M-1}\tan kd - 1}\right],
\end{gathered}
\end{equation}
and for $M$ odd
\begin{equation}
\begin{gathered}
\dfrac{q}{k}\tanh\frac{qd}{2} = \dfrac{\delta_M\tan kd - 1}{\delta_M + \tan kd},\\
\delta_{M, \mathrm{\ odd}} = \dfrac{k}{q}\left[\dfrac{\delta_{M-1} + \tanh qd}{\delta_{M-1}\tanh qd + 1}\right]
\end{gathered}
\end{equation}
and $\delta_1=0$.

But the solutions above are only for the $S(NS)_M$ case! The methodology is identical for $N(SN)_M$, and the results are, for $M$ even,
\begin{equation}
\begin{gathered}
\dfrac{q}{k}\tanh\frac{qd}{2} = \dfrac{\zeta_M\tan kd - 1}{\zeta_M + \tan kd},\\
\zeta_{M,\ \mathrm{even}} = \dfrac{k}{q}\bigg[\dfrac{\zeta_{M-1} + \tanh qd}{\zeta_{M-1}\tanh qd + 1}\bigg],
\end{gathered}
\end{equation}
and for $M$ odd
\begin{equation}
\begin{gathered}
\dfrac{k}{q}\tan\frac{kd}{2} = \dfrac{\zeta_M\tanh qd + 1}{\zeta_M + \tanh qd},\\
\zeta_{M,\ \mathrm{odd}} = \dfrac{q}{k}\bigg[\dfrac{\zeta_{M-1} + \tan kd}{1 - \zeta_{M-1}\tan kd}\bigg]
\end{gathered}
\end{equation}
and, again, $\zeta_1=0$.

\subsection{Neumann Condition}
The derivation for the Neumann condition eigenvalue equations is much like that of the Dirichlet condition. The only difference is, of course, the obvious change in the boundary condition---the Neumann condition necessitates that $\psi^\prime=0$ at the extreme boundaries. A derivation as detailed as the one for the Dirichlet condition would be superfluous here, as the methodology is exactly the same. But for added clarity, we will briefly consider the $SNS$ (one period) case. Here we find that the relevant wavefunctions come in the forms $\psi_0(x) = \alpha_0\cosh(qx)$ and $\psi_1(x) = \alpha_1\cos[k(\frac{3d}{2}-x)]$. (Notice the change in $\psi_1$ from the Dirichlet case: the sine term switched to a cosine so that $\psi'(\frac{3d}{2})=0$.) After we have mandated continuity of $\psi$ and $\psi^\prime$ at the internal interfaces, for $M=1$ we get $\frac{q}{k}\tanh\frac{qd}{2} = \tan kd$. Omitting the rest of the derivations, what we find is that for $M$ even,
\begin{equation}
\begin{gathered}
\dfrac{k}{q}\tan\frac{kd}{2} = \dfrac{\tanh qd + \gamma_M}{1 + \gamma_M\tanh qd}, \\
\gamma_{M, \mathrm{\ even}} = -\dfrac{k}{q} \bigg[\dfrac{\tan kd - \gamma_{M-1}}{1 + \gamma_{M-1}\tan kd}\bigg],
\label{SNS Neum even}
\end{gathered}
\end{equation}
and for $M$ odd,
\begin{equation}
\begin{gathered}
\dfrac{q}{k}\tanh\frac{qd}{2} = \dfrac{\tan kd - \gamma_M}{1 + \gamma_M\tan kd},\\
\gamma_{M, \mathrm{\ odd}} = \dfrac{q}{k}\bigg[\dfrac{\tanh qd + \gamma_{M-1}}{1 + \gamma_{M-1}\tanh qd}\bigg],
\label{SNS Neum odd}
\end{gathered}
\end{equation}
and $\gamma_1=0$.
For the $N(SN)_M$ case the procedure is no different. The recursive equations that emerge are, for M even,
\begin{equation}
\begin{gathered}
\dfrac{q}{k}\tanh\frac{qd}{2} = \dfrac{\tan kd - \varepsilon_M}{1 + \varepsilon_M\tan kd},\\
\varepsilon_{M,\ \mathrm{even}} = \dfrac{q}{k}\bigg[\dfrac{\tanh qd + \varepsilon_{M-1}}{1 + \varepsilon_{M-1}\tanh qd}\bigg],
\label{NSN Neum even}
\end{gathered}
\end{equation}
and for $M$ odd
\begin{equation}
\begin{gathered}
\dfrac{k}{q}\tan\frac{kd}{2} = \dfrac{\tanh qd + \varepsilon_M}{1 + \varepsilon_M\tanh qd},\\
\varepsilon_{M,\ \mathrm{odd}} = \dfrac{k}{q}\bigg[\dfrac{\varepsilon_{M-1} - \tan kd}{1 + \varepsilon_{M-1}\tan kd}\bigg]
\label{NSN Neum odd}
\end{gathered}
\end{equation}
and $\varepsilon_1=0$. The results in Eqs. \ref{SNS Neum even}--\ref{NSN Neum odd} actually replicate those obtained using the de Gennes-Werthamer approach,\cite{PRB Paper} as we should expect since the Neumann condition what is used in traditional solutions to these systems.

\section{Analysis}
Now that we have established the fundamental equations that describe the eigenvalues of our systems, we begin to look at and discuss the implications of the solutions.  Numerical values of $E(V)$ were obtained using the commercial program Mathematica\textsuperscript{TM}.
\begin{figure}
\includegraphics*[width=5.0in]{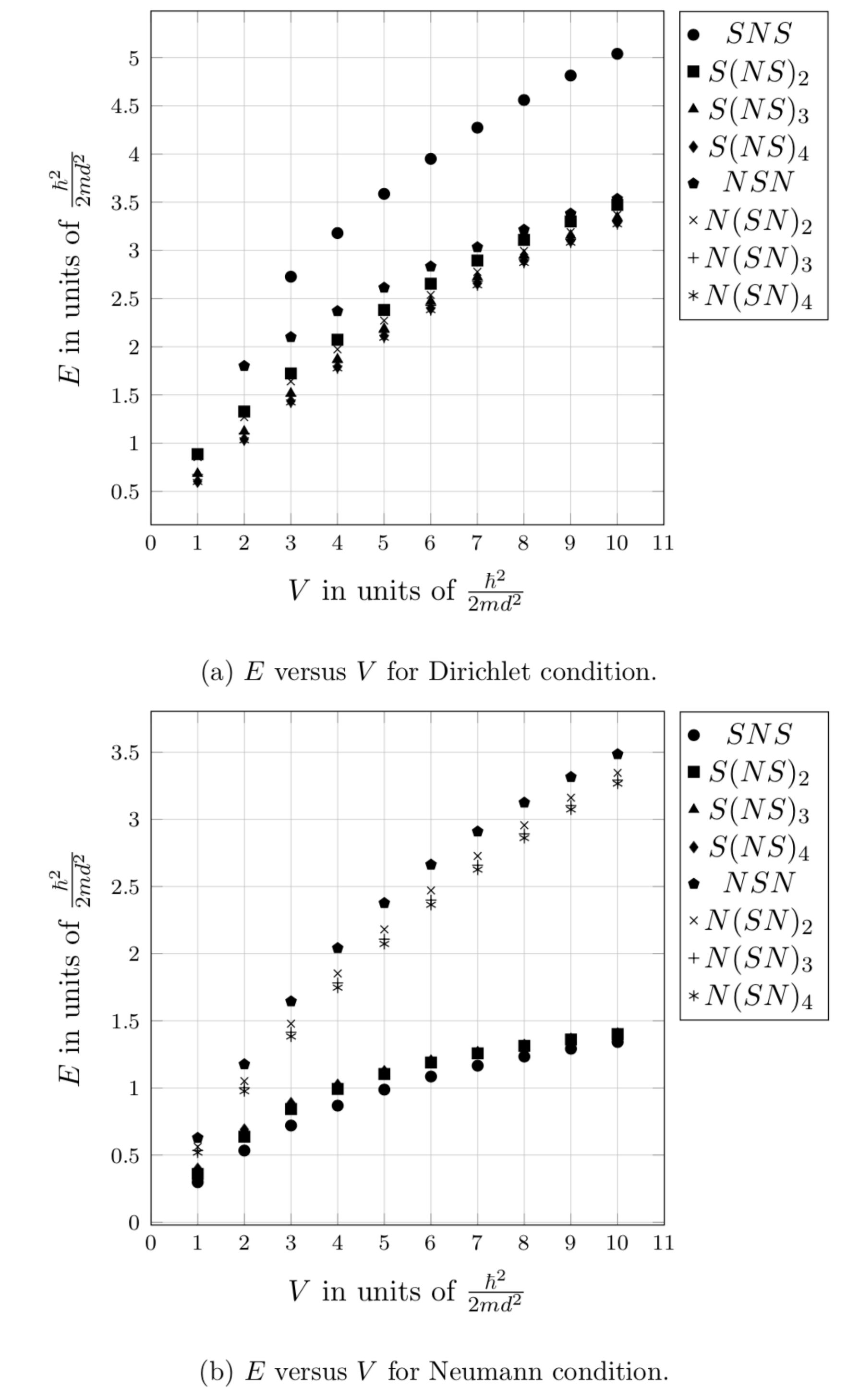}
\caption{Energy versus potential for Dirichlet and Neumann $S(NS)_M$ and $N(SN)_M$. Notice that the solutions for both $S(NS)_M$ and $N(SN)_M$ converge as $M$ increases for Dirichlet. Note that in the Dirichlet case, solutions where $E<V$ do not exist for $SNS$ when $V=1,2$, nor for $NSN$ when $V=1$.}
\label{EvV}
\end{figure}

\begin{figure}
\includegraphics*[width=7.0in]{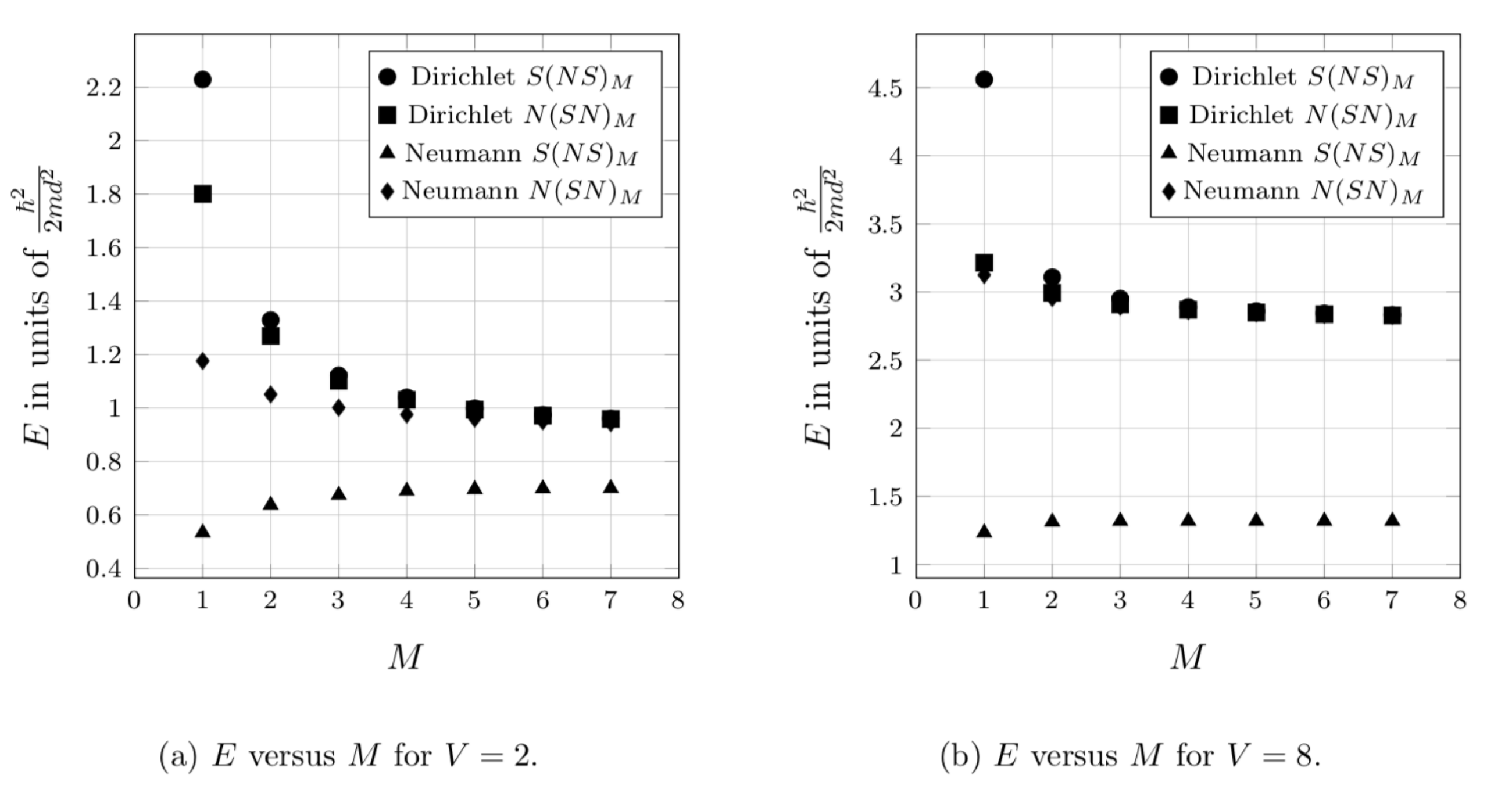}
\caption{Energy versus period for Dirichlet and Neumann $S(NS)_M$ and $N(SN)_M$ for (a) $V=2$ and (b) $V=8$. }
\label{EvM}
\end{figure}
The Dirichlet solutions behave in the expected way. What one first notices from Fig. \ref{EvV}a is that the $N(SN)_M$ and $S(NS)_M$ solutions converge with increasing $V$ (with the exception of $M=1$), and with increasing $M$, the convergence is even more pronounced across all $V$. In terms of a proximity effect system, this translates to a convergence of $T_c$ as the layer width $d$ goes down. A check was also done to verify the behavior of this system at low $V$, as we would expect a linear regression of the eigenvalues as $V\rightarrow 0$, as well as convergence of like-period solutions for $S(NS)_M$ and $N(SN)_M$ systems.\footnote{This is because of perturbation theory, which states that small changes in $E$ for $V$ near $V=0$ can be described by $\delta E = \langle\psi|V|\psi\rangle$.} This behavior was confirmed.

On the other hand, the Neumann solutions show different results. In Fig. \ref{EvV}b, the $S(NS)_M$ and $N(SN)_M$ solutions take starkly different paths as $V$ increases. Within $S(NS)_M$ or $N(SN)_M$ categories the eigenvalues do appear to converge as $V$ goes up, but we would like to see solutions between categories converge, and we do not.

The most important piece of information here, however, is the behavior of $E$ as $M$ increases as shown in Fig. \ref{EvM} for $V=2$ and $V=8$. In the Dirichlet case, it is obvious that solutions of $S(NS)_M$ and $N(SN)_M$ converge for increasing $M$. This is exactly what we would expect---the larger the systems get, the more similarly they behave. However, the Neumann case does not behave this way. The Neumann case seems to have a dependence upon the end layers (i.e., whether the system is contained by $S$ layers or $N$ layers). The failure of these systems to converge is precisely what is obtained using the standard de Gennes-Werthamer approach (i.e., using Eq. \ref{prox eq}).\cite{PRB Paper} Interestingly, the Neumann $N(SN)_M$ case does show convergence with the Dirichlet solutions as $M$ increases---it is the $S(NS)_M$ case that remains separate from the other solutions.

\begin{figure}
\centering
\includegraphics*[width=5.0in]{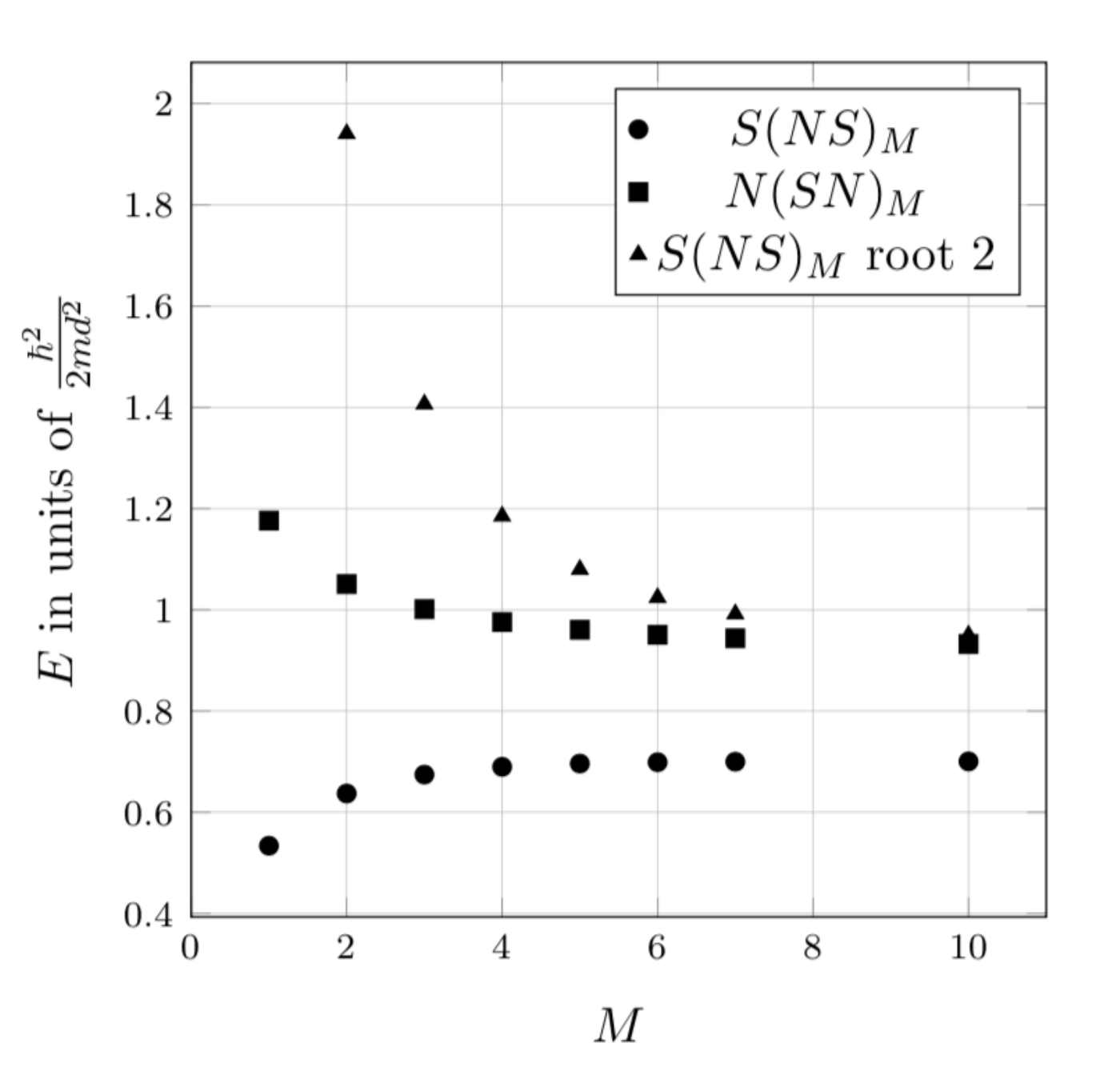}
\caption{Energy versus period with the second solution of the Neumann $S(NS)_M$ scenario. Here we have to go out to $M=10$ to be sure of the convergence. There does not exist an eigenvalue for $S(NS)_M$ root 2 for $M=1$ such that $E<V$.}
\label{EvM Neum 2}
\end{figure}

However, despite the ground state eigenvalues of $S(NS)_M$ Neumann not converging with those of $N(SN)_M$ Neumann, the second root of the solution set for $S(NS)_M$ Neumann did demonstrate convergence with the Neumann $N(SN)_M$ (and likewise the Dirichlet solutions). These results are shown in Fig. \ref{EvM Neum 2}. Unfortunately, the second root solutions of the $S(NS)_M$ set are not viable solutions because the corresponding wavefunctions have nodes, which is a violation of the requirement that $\psi(x)$ must be positive definite. Incidentally, this same pattern occurs for eigenvalues corresponding to odd solutions of $\psi(x)$ for Schr\"odinger's equation, which, of course, are not allowed for the same reason.

\begin{figure}
\includegraphics*[width=7.0in]{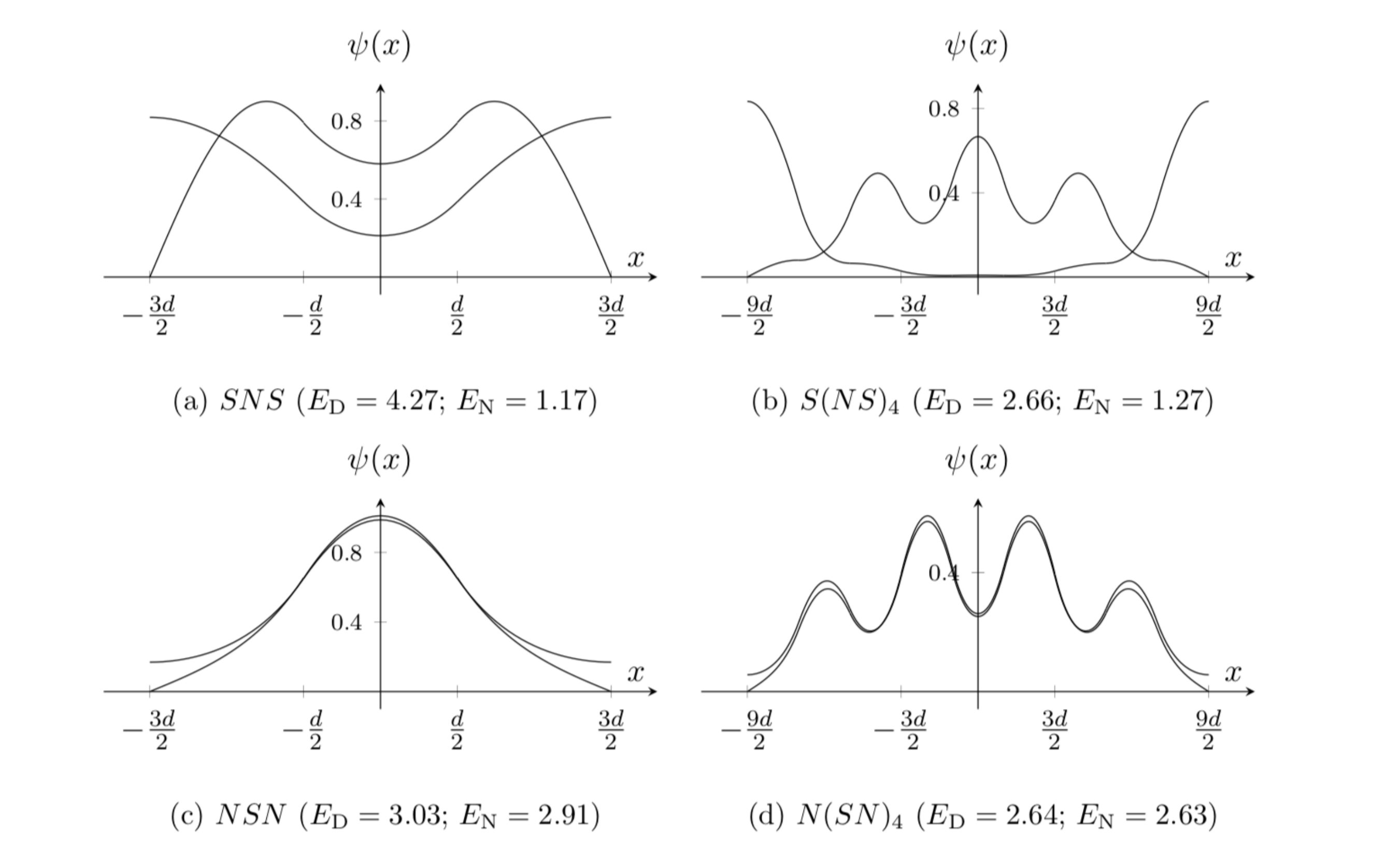}
\caption{Sample of wavefunctions for periods $M=1,4$ of both the $S(NS)_M$ and $N(SN)_M$ systems under both Dirichlet and Neumann conditions. The Dirichlet condition is indicated by black and the Neumann by blue. Potential is $V=7$ for all wavefunctions, and all wavefunctions are normalized.}
\label{Wavefunctions}
\end{figure}

Finally, a sample of the resulting wavefunctions is shown in Fig. \ref{Wavefunctions}. All of the wavefunctions are normalized, although the proximity effect does not require it (``normalization" of $\Delta(x)$ in traditionally solved systems would be determined by measured properties of the sample). As expected, we see that the wavefunctions put the bulk of the probability in the $S$ layers.

\section{Conclusion}
What the results show is that there is a strong dependence between the boundary condition and the convergence of the eigenvalues. As we saw, the Dirichlet case resulted in the convergence of the ground state eigenvalues while the Neumann case did not. In terms of a proximity effect system, this would mean that $T_c$ converges as $M\rightarrow\infty$ in the Dirichlet case and not in the Neumann case. While this is seemingly a very strange theoretical result, this is precisely what one finds when using traditional methods to solve proximity effect systems, as the Neumann condition is a physical requirement.\cite{PRB Paper} The fact that convergence was identified between the second root of the $S(NS)_M$ Neumann solutions and the ground state $N(SN)_M$ Neumann certainly opens more questions as to what exactly this might mean, especially in light of the requirements for $\psi(x)$. More work ought to be done to investigate the nature and behavior of the $N(SN)_M$ Neumann solutions to see why the ground state does not converge with the $S(NS)_M$ Neumann (and also Dirichlet) solutions.\footnote{The data that support the findings of this study are available from the corresponding author upon reasonable request.}

\end{document}